\documentclass[reprint,showpacs,preprintnumbers,pre,9pt]{revtex4-1}
\usepackage{graphicx}
\begin{document}\small

\title{Recent theoretical advances in elasticity of membranes following Helfrich's spontaneous curvature model}
\author{Z. C. Tu}\email[Email: ]{tuzc@bnu.edu.cn}
\affiliation{Department of Physics, Beijing Normal University, Beijing 100875, China}

\author{Z. C. Ou-Yang}\email[Email: ]{oy@itp.ac.cn}
\affiliation{Institute of Theoretical Physics, Chinese Academy of
Sciences, Beijing 100080, China}

\begin{abstract}
Recent theoretical advances in elasticity of membranes following Helfrich's famous spontaneous curvature model are summarized in this review. The governing equations describing equilibrium configurations of lipid vesicles, lipid membranes with free edges, and chiral lipid membranes are presented. Several analytic solutions to these equations and their corresponding configurations are demonstrated.

\preprint{Adv. Colloid. Interface Sci. 208 (2014) 66-75\hspace{3cm}}
\end{abstract}\startpage{66}


\maketitle

\section{Introduction}
Cells are basic elements of life. Membrane structures make cells to be relatively independent individuals but still able to exchange matter and energy between the inner sides of cells and the outer surroundings. Lipids and proteins are the main chemical components of membranes. Under the physiological condition, they form a fluid mosaic structure~\cite{nicolson}. In this fluid mosaic model, a cell membrane is considered as a lipid bilayer where lipid molecules can move freely in the membrane surface like
fluid, while proteins are embedded in the lipid bilayer. Although the cell membrane possesses the character of fluid membranes, it is not a fully 3-dimensional (3D) isotropic fluid. Actually, the cell membrane is in the liquid crystal phase~\cite{helfrich73} and it can endure the out-of-plane deformation of bending. This physical property is crucial to the morphology and function of cells.

The simplest cells are human red blood
cells in mature stage because they have no internal organelles. Therefore, the physical property of membranes uniquely determines the shapes of red blood cells.
Normal human red blood cells at rest are typically of biconcave discoidal shape, so we call them discocytes. Why are red blood cells at rest always of biconcave discoidal shape? This problem has attracted
considerable attention of researchers. In order to fit the biconcave shape, Fung and Tong~\cite{Fung68} proposed a sandwich model with an assumption that the thickness of the membrane could vary in the scale of micrometers. However,
the observation with an electron microscope revealed that the thickness of the membrane
should be uniform in the scale of micrometers~\cite{Pinder72}. Lopez \emph{et al.}~\cite{Lopez68} proposed that the distribution of electric
charges over membrane surface might vary in the scale of micrometers. However, the measurement by Greer and Baker~\cite{Greer70}
revealed a uniform distribution of charges over the scale of micrometers in the surface of red blood cells.
Murphy~\cite{Murphy65} argued that the shape of red blood cells might be related to the nonuniform distribution of cholesterol in the cell membrane. But the experiment by Seeman \emph{et al.}~\cite{Seeman73} did not
support this hypothesis.
Canham~\cite{Canham70} proposed an incompressible shell model and argued that the biconcave discoidal shape should be the result of minimizing the curvature energy for the given surface area and volume of the red blood cell. Although the dumbbell-like shape has the
same curvature energy as the biconcave discoid within this model, the former configuration has never been observed in any experiment~\cite{Helfrich75}.

Helfrich recognized that a lipid bilayer, the main ingredient of cell membranes, is in the liquid crystal state~\cite{helfrich73}. A membrane is thought of as a 2D smooth surface in a 3D Euclidean space because its thickness is much smaller than its lateral dimension. {By analogy with the Frank energy~\cite{Frank58} of a bent nematic crystal box}, Helfrich derived the curvature energy per unit area of the membrane~\cite{helfrich73}:
\begin{equation}\mathcal{E}_\mathrm{H}=(k_c/2) (2H +c_0)^2 + \bar{k} K,\label{eq-helfrichcuv}
\end{equation}
where $k_c$ and $\bar{k}$ are two bending moduli. {The measured value of $k_c$ is about tens of $k_BT$, the energy scale of thermal motion, and it depends on the constituents of lipid bilayer~\cite{Nagle2013,Nagle2008,NagleBJ08,NagleBJ06,SaldittPRL04,SorrePNAS09,TianBJ09}.} There is still a lack of directly experimental schemes to extract the value of $\bar{k}$. $H$ and $K$ in equation~(\ref{eq-helfrichcuv}) are locally the mean curvature and the Gaussian curvature of the membrane surface, respectively. The parameter $c_0$ in equation~(\ref{eq-helfrichcuv}) is called spontaneous curvature which reflects the asymmetry between two leaves of the membrane. {The Canham curvature energy can be regarded as the special form of the Helfrich curvature energy with $c_0=0$ and $\bar{k}=-k_c$. Although the Helfrich curvature energy (\ref{eq-helfrichcuv}) was originally derived from the liquid crystal theory, it can be utilized to describe the bending energy of 2D isotropic membranes. Since a normal red blood cell has no internal organelles, it can be regarded as an amount of liquid enclosed by a cell membrane. The cell membrane consists of not only a lipid bilayer but also a layer of membrane skeleton beneath the lipid bilayer~\cite{Sackmannbook}. The lipid bilayer is 2D isotropic. The membrane skeleton of red blood cell is roughly a 2D hexagonal lattice. As
is well known, the mechanical property of a 2D hexagonal lattice is the same as that of
2D isotropic materials~\cite{nyebook}. Thus, the cell membrane of red blood cell, a lipid bilayer plus a layer of membrane skeleton, may be regarded as locally 2D isotropic matter so that its bending energy up to the quadratic order of curvatures may still be expressed as the Helfrich curvature energy.}
The equilibrium shape of a closed membrane is thought of as the configuration minimizing the total Helfrich curvature energy $\int \mathcal{E}_\mathrm{H} \mathrm{d}A$ for the given surface area of the membrane and volume enclosed in the membrane. With the consideration of equation~(\ref{eq-helfrichcuv}), {both numerical and theoretical results could be achieved~\cite{Helfrich76,NaitoPRE93} to fit the biconcave discoidal shape of human red blood cells.}

Following Helfrich's spontaneous curvature model, the elasticity of membranes has been deeply investigated in the past forty years~\cite{LipowskyN91,Seifert97ap,OYbook1999,Mladenovctp13}. In this review, we will report the most relevant theoretical advances following the Helfrich model in terms of our prospect. In section~\ref{sec-lipves}, we briefly present the shape equation of lipid vesicles and its {special solutions}. The nonlocal bending theory, a generalization of Helfrich's model, is analyzed. In section~\ref{sec-OpenLBfdg}, we present the governing equations describing equilibrium configurations of lipid membranes with free edges and the corresponding {special solutions}. The stress tensor of fluid membranes and the values of Gaussian bending modulus $\bar{k}$ in equation~(\ref{eq-helfrichcuv}) are discussed. In section~\ref{sec-CLM}, the Helfrich model is generalized to the theory of chiral lipid membranes. With the consideration of a concise theory of chiral lipid membranes, the governing equations describing equilibrium configurations of chiral lipid membranes with or without free edges and the corresponding special solutions are demonstrated. The last section is a brief summary where we also propose several theoretical challenges in the elasticity of membranes.

\section{Lipid vesicles\label{sec-lipves}}
Lipid molecules are amphiphilic. When a certain amount of lipid molecules are dispersed in water, they may self-organize into vesicles with different configurations. In this section, we will present the theoretical work relevant to the configurations of lipid vesicles.
\subsection{Shape equation and its special solutions\label{sec-Lipshpcls}}
A lipid vesicle can be regarded as a closed smooth surface. Its equilibrium configuration is expected to correspond to the local minimal of the extended Helfrich's free energy
\begin{equation}\label{eq-exthelfrich}
F_\mathrm{H}=\int\mathcal{E}_\mathrm{H}\mathrm{d}A + \lambda A+ pV,
\end{equation}
where $\mathcal{E}_\mathrm{H}$ is the Helfrich curvature energy~(\ref{eq-helfrichcuv}). The symbols $A$ and $V$ represent the total area of the membrane surface and the volume enclosed in the vesicle, respectively. $\lambda $ and $p$ are two Lagrange multipliers which constrain the fixed $A$ and $V$, respectively. These two parameters can be regarded as the apparent surface tension and osmotic pressure (the outside pressure minus the inside one) of the lipid vesicle, respectively.

The Euler-Lagrange equation corresponding to functional (\ref{eq-exthelfrich}) may be derived from the calculus of variation. {The first order variation of functional (\ref{eq-exthelfrich}) without $c_0$ was calculated by Jenkins~\cite{Jenkins77}.} With the consideration of the spontaneous curvature $c_0$, Ou-Yang and Helfrich~\cite{OYPRL87,OYPRA87} obtained the general Euler-Lagrange equation:
\begin{equation}\tilde{p}-2\tilde{\lambda}
H+(2H+c_0)(2H^2-c_0H-2K)+\nabla^2(2H)=0,\label{eq-shapeclosed}\end{equation}
with a reduced osmotic pressure $\tilde{p}\equiv p/k_c$ and a reduced surface tension $\tilde{\lambda}\equiv \lambda/k_c$. $\nabla^2$ is the Laplace operator defined on a 2D surface.
This formula is called shape equation of lipid vesicles, which represents the force balance along the normal direction of the membrane surface. If we use $f(x,y,z)=0$, the function of spatial coordinates $x$, $y$ and $z$, to express the surface, equation~(\ref{eq-shapeclosed}) is a fourth-order nonlinear partial differential equation. There is no general solution to a nonlinear differential equation, so we can only guess some special
solutions in terms of intuition.

\begin{figure}[pth!]
\centerline{\includegraphics[width=6cm]{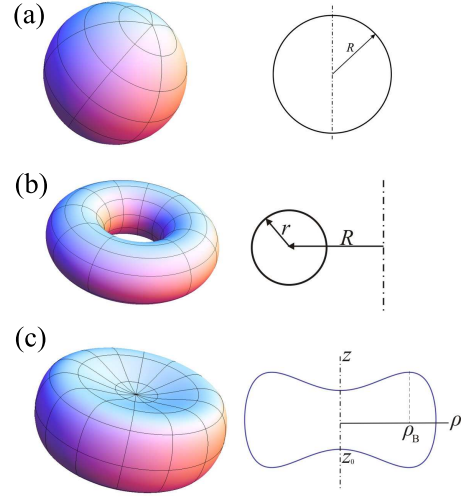}}
\caption{(Color online) Special
solutions to shape equation (\ref{eq-shapeclosed}) and their generation curves: (a) Sphere; (b) Torus; (c) Biconcave discoid.\label{fig-vesicles}}
\end{figure}

\begin{figure}[pth!]
\centerline{\includegraphics[width=6cm]{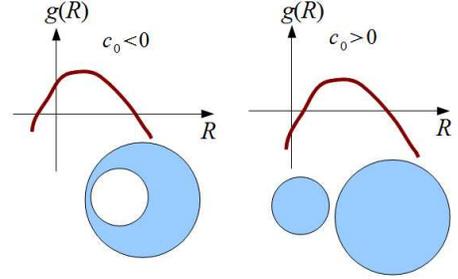}}
\caption{(Color online) Diagram of function $g(R)\equiv\tilde{p}R^2+(2\tilde{\lambda}+c_0^2)
R-2c_0$ and roots to equation~(\ref{sphericalbilayer}). There exist two roots when $\tilde{p}<0$ and $(2\tilde{\lambda}+c_0^2)^2+8\tilde{p}c_0>0$.\label{fig-endocytosis}}
\end{figure}

The simplest solution is a spherical surface with radius $R$ as shown in figure~\ref{fig-vesicles}a. In this case, shape equation (\ref{eq-shapeclosed}) requires
\begin{equation}\tilde{p}R^2+2\tilde{\lambda}
R-c_0(2-c_0R)=0.\label{sphericalbilayer}\end{equation}
The existence of roots to the above equation depends on the values of parameters $c_0$, $\tilde{p}$, and $\tilde{\lambda}$. In particular, when $\tilde{p}<0$ and $(2\tilde{\lambda}+c_0^2)^2+8\tilde{p}c_0>0$, there are two roots to equation~(\ref{sphericalbilayer}) as shown in figure~\ref{fig-endocytosis}, {which suggests that we might observe the coexistence of lipid vesicles with difference radii in a solution of lipid molecules.}

The spherical vesicle is instable when the reduced osmotic pressure $\tilde{p}$ exceeds a threshold
$\tilde{p}_{2}\equiv 2(6-c_0R)/R^3$ and it will be transformed into a biconcave discoid~\cite{helfrich73}. When the reduced osmotic pressure exceeds $\tilde{p}_{l}\equiv 2[l(l+1)-c_0R]/R^3$, the vesicle will be further transformed into a shape of $l$-th polygon symmetry~\cite{OYPRL87}.

The second solution to shape equation (\ref{eq-shapeclosed}) is a torus with the ratio of two generation radii being $\sqrt{2}$~\cite{oypra90}. As shown in figure~\ref{fig-vesicles}b, the
torus can be expressed as a vector form
$\{(R+r\cos\varphi)\cos\theta,(R+r\cos\varphi)\sin\theta,r\sin\varphi\}$.
Through simple calculations, shape equation (\ref{eq-shapeclosed}) is transformed into
\begin{eqnarray}
&&\hspace{0.26cm} [(2 c_0^2r^2-4 c_0r+4\tilde{\lambda}
r^2+2\tilde{p}r^3)/{\varrho^3}]\cos^3\varphi\nonumber\\
&&+[(5 c_0^2r^2-8 c_0r+10\tilde{\lambda} r^2+6\tilde{p}r^3)/{\varrho^2}]\cos^2\varphi\nonumber\\
&&+[{(4 c_0^2r^2-4 c_0r+8\tilde{\lambda} r^2+6\tilde{p}r^3)}/{\varrho}]\cos\varphi\nonumber\\
&&+2/{\varrho^2}+(c_0^2r^2-1)+2(\tilde{p}r+\tilde{\lambda})r^2=0\label{eq-torus}
\end{eqnarray}
with $\varrho\equiv R/r$. This equation holds only if the coefficients of
$\{1,\cos\varphi,\cos^2\varphi,\cos^3\varphi\}$ vanish for finite $\varrho$, which follows $\tilde{p}=-2c_0/r^2$, $\tilde{\lambda}=c_0(4-c_0r)/2r$, and
$\varrho\equiv R/r=\sqrt{2}$.
That is, there exists a lipid torus with the ratio of its two
generation radii being $\sqrt{2}$, which was confirmed by the
experiment~\cite{MutzPRA91}. {We emphasize that the torus with $\varrho\equiv R/r=\sqrt{2}$ is not only a solution to shape equation (\ref{eq-shapeclosed}), but also a solution to the Willmore equation ($\nabla^2 H-2KH+2H^3=0$) which is the Euler-Lagrange equation corresponding to the functional $\int H^2 \mathrm{d}A$. In 1965, Willmore~\cite{Willmore1965} conjectured that the integral of the square of the mean curvature of a smooth immersed surface with toroidal topology in the 3D Euclid space is at least $2\pi^2$ and the lower bound is taken for the torus with $\varrho\equiv R/r=\sqrt{2}$. This conjecture has been investigated by many mathematicians~\cite{LiYau,Bryant84,Simon93,Topping20,BauerKuwert} for fifty years. Finally, Marques and Neves succeeded in proving this conjecture by using the min-max theory of minimal surfaces~\cite{MNeves14}.}

The third solution to shape equation (\ref{eq-shapeclosed}) corresponds to a biconcave discoid shown in figure~\ref{fig-vesicles}c. To make it clear, we consider the axisymmetric form of shape equation (\ref{eq-shapeclosed}). Since each axisymmetric surface may be generated by a contour line, we merely need to derive out the governing equation for the contour line. As proposed by Helfrich~\cite{helfrich73}, the contour line can be parameterized as function $\psi=\psi(\rho)$ where $\rho$ represents the rotation radius of some point in the contour line while $\psi$ is the tangent angle of the contour line at that point. In this representation, shape equation (\ref{eq-shapeclosed}) is transformed into~\cite{HuOY93PRE}:
\begin{equation}
\tilde{p}+\tilde{\lambda}
\mathcal{H}+(c_{0}-\mathcal{H})\left(\frac{\mathcal{H}^{2}}{2}+\frac{c_{0}\mathcal{H}}{2}-2K\right)-\frac{\cos \psi }{\rho}(\rho\cos \psi \mathcal{H}')'=0,\label{eqshapesymmetr}\end{equation}
with $\mathcal{H}\equiv {\sin \psi }/{\rho}+(\sin\psi)'$ and $K={\sin \psi
}(\sin\psi)'/{\rho}$. The `prime' represents the derivative with respect to $\rho$.
This equation is a third-order ordinary differential
equation. {Zheng and Liu found a first integral $\eta_{0}$ for the above equation~\cite{zhengliu93} and then transformed this equation
into a second-order differential equation:}
\begin{equation}\cos\psi \mathcal{H}'
+(\mathcal{H}-c_{0}) \sin\psi\psi^{\prime}-\tilde{\lambda} \tan\psi+\frac{\eta_{0}/\rho-\tilde{p}\rho/2}{\cos\psi}-\frac{\tan\psi}
{2}(\mathcal{H}-c_{0})^{2} =0.\label{firstintg}\end{equation}
{Castro-Villarreal and Guven~\cite{GuvenJPA07} pointed out that the existence of the first integral results from the conservation law of the stress in fluid membranes.} The above equation degenerates into the formula derived by Seifert \emph{et al.} \cite{SBLPRA91} when $\eta_{0}=0$ in equation~(\ref{firstintg}) which holds for vesicles with spherical topology free from singular points \cite{Podgornikpre95}.

When $0<c_0\rho_B<\mathrm{e}$, the parametric equation
\begin{equation}\left\{\begin{array}{l}\sin\psi=-c_0\rho\ln(\rho/\rho_B)\\
z=z_0+\int_0^\rho \tan\psi \mathrm{d}\rho
\end{array}\right.\label{solutionbicon}\end{equation}
corresponds to a contour line shown in figure~\ref{fig-vesicles}c. Substituting it into equation~(\ref{firstintg}),
one obtains $\tilde{p}=0$, $\tilde{\lambda}=0$, and
$\eta_0=-2c_0 \neq 0$. That is, a biconcave
discoid generated by the contour line satisfying equation~(\ref{solutionbicon}) is a special solution to the shape equation of vesicles. {It is found that this special solution can fit the biconcave discoidal shape of human red blood cells under normal physiological
conditions~\cite{NaitoPRE93,NaitoPRE96}.}

There also exist several analytic solutions to shape equation (\ref{eq-shapeclosed}), such as the constant-mean-curvature surfaces (excluding spheres) and cylinder-like surfaces~\cite{NaitoPRL95,Konop97,MladenovEPJB02,GuvenPRE20022D,Castro07,MladenovJPA08,ZhangOYPRE96,Djondjorov10}. We have not explicitly sketched them because they do not correspond to closed configurations without self-intersections.

\subsection{Generalization: nonlocal bending theory}
To explain the stomatocyte-discocyte-echinocyte transition of human red blood cells, the Helfrich model is generalized to a nonlocal bending theory. When a lipid bilayer is bent from a flat configuration, the area of per lipid molecule in each leaf should depart from the equilibrium value. Considering the in-plane stretching or compression in each leaf, a nonlocal term $(k_r/2)(\int 2 H \mathrm{d}A)^2$ may be added to the bending energy of membranes~\cite{Evans1980,Svetina1985}. $k_r =k_a t^2/2A_0$ is an elastic constant where $k_a$ and $t$ being the compression modulus and thickness of the monolayer, respectively. $A_0$ is the prescribed area of the membrane in the flat configuration. Considering this term, one may express the free energy of a vesicle as
\begin{equation}\label{eq-feADE1}
F_\mathrm{NL}=\int \mathcal{E}_\mathrm{H}\mathrm{d}A + \lambda A+ pV+\frac{k_r}{2}\left(\int 2H \mathrm{d}A\right)^2.
\end{equation}
This model is called bilayer-coupled model~\cite{Evans1980,Svetina1985}.

Similarly, if the membrane is initially curved with (spontaneous) relative area difference $a_0$, the nonlocal term $(k_r/2)(\int 2H \mathrm{d}A+a_0)^2$ may be included in the free energy after the membrane is deformed~\cite{miaoling1996}.
Thus the free energy of a vesicle may be expressed as
\begin{equation}\label{eq-feADE2}
F_\mathrm{ADE}=\int \mathcal{E}_\mathrm{H} \mathrm{d}A + \lambda A+ pV+\frac{k_r}{2}\left(\int 2H \mathrm{d}A+a_0\right)^2.
\end{equation}
This model is called area-difference-elasticity model~\cite{miaoling1996}. Based on this model and numerical simulations, Lim et al.~\cite{Lim2002} explained the stomatocyte-discocyte-echinocyte transition of human red blood cells. The budding transitions of axisymmetric fluid-bilayer vesicles have been fully investigated on the basis of area difference elasticity~\cite{miaoling1996}. It is still necessary to discuss the general cases without the presumption of axisymmetry.

In fact, if we make variable transformations $C_0=c_0+a_0k_r/k_c$ and $\Lambda=\lambda+a_0^2k_r/2A_0-k_ra_cc_0-k_r^2a_0^2/2k_c$, the above free energy is transformed into the form of equation (\ref{eq-feADE1}). Thus it is sufficient for us to consider the free energy (\ref{eq-feADE1}).
According to the variational method developed in the previous work~\cite{TuJPA04}, the shape equation of vesicles which corresponds to the Euler-Lagrange equation of free energy (\ref{eq-feADE1}) can be derived as
\begin{equation}\tilde{p}-2\tilde{\lambda}
H+(2H+c_0)(2H^2-c_0H-2K)+\nabla^2(2H) -4\tilde{k}_rK\int H \mathrm{d}A=0\label{shape-adeclosed}\end{equation}
with reduced parameters $\tilde{p}\equiv p/k_c$, $\tilde{\lambda}\equiv \lambda/k_c$ and $\tilde{k}_r\equiv k_r/k_c$. This is a fourth-order nonlinearly integro-differential equation, so it is very difficult for us to seek analytic solutions to this equation.

Let us assume $c_0 = \bar{c}_0-2\tilde{k}_r\int H\mathrm{d}A$, then equation (\ref{shape-adeclosed}) is transformed into
\begin{equation}\tilde{p}-2\bar{\lambda}
H+(2H+\bar{c}_0)(2H^2-\bar{c}_0H-2K)+\nabla^2(2H)=0\label{shp-adeclosedtran}\end{equation}
with $\bar{\lambda}=\tilde{\lambda}+(c_0^2-\bar{c}_0^2)/2$. Since the above equation has the same form as shape equation (\ref{eq-shapeclosed}), the solutions to shape equation (\ref{shape-adeclosed}) should share the same forms as those to shape equation (\ref{eq-shapeclosed}) except for different values of parameters.

\section{Lipid membranes with free edges\label{sec-OpenLBfdg}}
{Open bilayer configurations can be stabilized by edge-reactant salts~\cite{Fromherz83,Fromherz86} or some proteins~\cite{Hotani98}. This experimental fact draws researchers' attention to studying the configurations of lipid membranes with free exposed edges. Baol and Rao~\cite{Boal92} found that the only energy minimizing axisymmetric shapes are the disk and the sphere for zero spontaneous curvature, and that the transition from open to closed configuration depends on the rigidity and the line tension. Capovilla \emph{et al.} investigated the stress of fluid membranes~\cite{Capovilla2} and then derived the general governing equations of lipid membranes with free exposed edges~\cite{Capovilla}. An equivalent form of the general governing equations was also derived from variational method with aid of differential forms~\cite{TuPRE03}. The possible solutions to the governing equations of lipid membranes with free edges were discussed in recent work~\cite{TuJCP2010,Tucpb2013}.} We will sketch these results and their implications based on Hefrich's model in this section.

\subsection{Governing equations and their special solutions\label{sec-openLBs}}
\begin{figure}[pth!]
\centerline{\includegraphics[width=5cm]{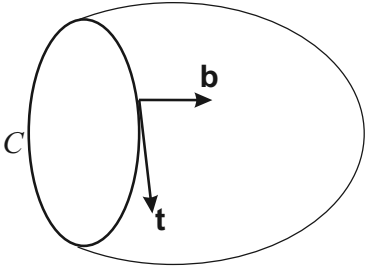}}
\caption{Open membrane is regarded as an open smooth surface with a boundary curve. \label{figopenm}}
\end{figure}

A lipid membrane with a free edge can be regarded as a smooth
surface with a boundary curve $C$ as shown in figure~\ref{figopenm}. Vectors $\mathbf{t}$ and $\mathbf{b}$ are located in the tangent plane of the surface. The former is the tangent vector of $C$ while the latter is perpendicular to $\mathbf{t}$ and points to the side that the surface is located in. Since the freely exposed edge is energetically unfavorable, we assign a positive line tension $\gamma$, the energy cost per unit length, to the free edge. Then the total free energy can be expressed as
\begin{equation}F_\mathrm{OM}= \int \mathcal{E}_\mathrm{H} \mathrm{d} A +\lambda A +\gamma L ,\label{eq-frenergyn2}\end{equation}
where $L$ is the total length of the free edge.

By using the moving frame method to calculate the first-order variation of functional (\ref{eq-frenergyn2}), Tu and Ou-Yang~\cite{TuPRE03} derived the shape equation
\begin{equation}(2H+c_{0})(2H^{2}-c_{0}H-2K)-2\tilde\lambda H+\nabla
^{2}(2H) =0, \label{eq-openm}\end{equation}
and three boundary conditions
\begin{eqnarray}
&&\left[ (2H+c_{0})+\tilde{k}\kappa_n\right]_{C} =0,\label{bound1} \\
&&\left[ -2{\partial H}/{\partial\mathbf{b}}+\tilde\gamma
\kappa_n+\tilde{k}\dot{\tau}_g\right] _{C} =0,\label{bound2}\\
&&\left[ (1/{2})(2H+c_{0})^{2}+\tilde{k}K+\tilde\lambda
+\tilde\gamma \kappa_{g}\right]_{C}=0\label{bound3},
\end{eqnarray}
where $\kappa_n$, $\kappa_g$, and $\tau_g$ are the normal
curvature, the geodesic curvature, and the geodesic torsion of the boundary curve, respectively. The `dot' represents the derivative with respect to the arc length of the edge. $\tilde{k}\equiv\bar{k}/k_c$ and $\tilde{\gamma}\equiv\gamma/k_c$ are the reduced bending modulus and the reduced line tension, respectively. According to the physical meaning of variation,
equation~(\ref{eq-openm}) indicates
the force balance in the normal direction of the membrane while equations (\ref{bound1})--(\ref{bound3}) represent the force or moment balances at each point in
curve $C$~\cite{TuPRE03}. Thus the above governing
equations are also available for an open membrane with several edges.

Since the points in the boundary curve should satisfy not only the
boundary conditions, but also the shape equation, above governing equations (\ref{eq-openm})--(\ref{bound3}) might not be independent of each other. In other words, there exist compatibility conditions for these equations. By using scaling transformation, Tu~\cite{TuJCP2010} derived a compatibility condition
\begin{equation}2 c_0 \int  H \mathrm{d} A+(2\tilde\lambda+c_0^2) A +\tilde\gamma L=0.\label{constraitg}\end{equation}

Through similar discussions in section~\ref{sec-Lipshpcls}, shape equation~(\ref{eq-openm}) may be transformed into a second-order ordinary differential equation
\begin{equation}\cos\psi \mathcal{H}'
+(\mathcal{H}-c_{0}) \sin\psi\psi^{\prime}-\tilde{\lambda} \tan\psi+\frac{\eta_{0}}{\rho\cos\psi}-\frac{\tan\psi}
{2}(\mathcal{H}-c_{0})^{2} =0\label{firstintg2}\end{equation} for an axisymmetric surface.
Comparing this equation with boundary conditions (\ref{bound1})--(\ref{bound3}), Tu~\cite{TuJCP2010} achieved another compatibility condition,
\begin{equation}\eta_{0}=0,\label{compat-cond}\end{equation}
for axisymmetric surfaces.
With the consideration of this condition, the shape equation
is reduced to
\begin{equation}\cos\psi \mathcal{H}'
+(\mathcal{H}-c_{0}) \sin\psi\psi^{\prime}
  -\tilde{\lambda} \tan\psi-\frac{\tan\psi}%
{2}(\mathcal{H}-c_{0})^{2} =0,\label{newshapeq}\end{equation}
while three boundary conditions are reduced to two independent equations as follows~\cite{TuPRE03,TuJCP2010}:
\begin{eqnarray}
\left[\mathcal{H}-c_{0}+\tilde{k}{\sin \psi }/{\rho}\right]_C=0,\label{sbound1}\\
\left[\frac{1}{2}(\mathcal{H}-c_0)^2+\tilde{k}K+\tilde{\lambda}-\sigma\tilde{\gamma}
\frac{\cos \psi }{\rho}\right]_C=0,\label{sbound3}
\end{eqnarray}
where $\sigma =1$ or $-1$ if the tangent vector $\mathbf{t}$ of the boundary curve is parallel or antiparallel to the rotation direction, respectively.

\begin{figure}[pth!]
\centerline{\includegraphics[width=7cm]{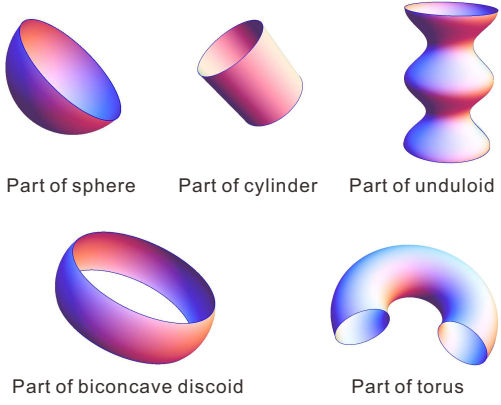}}
\caption{\label{fig-impopm}(Color online)
Schematics of several impossible open membranes with free edges.}
\end{figure}

An obviously but trivially analytic solution to shape equation~(\ref{eq-openm}) with boundary conditions~(\ref{bound1})--(\ref{bound3}) is a flat circular disk with radius $R$. In this case, equations~(\ref{eq-openm})--(\ref{bound3}) degenerate into
\begin{equation}\tilde{\lambda} R+ \tilde{\gamma} =0\end{equation} for vanishing $c_0$.
It is a stiff task to find nontrivially analytic solutions to governing equations~(\ref{eq-openm})--(\ref{bound3}). To do that, we need to seek a surface satisfying shape equation~(\ref{eq-openm}), and then find a simple closed curve abiding by boundary conditions~(\ref{bound1})--(\ref{bound3}) on this surface. Through a sophisticated analysis, Tu~\cite{TuJCP2010,Tucpb2013} proved a theorem of non-existence: For finite line tension, there does NOT exist an open membrane being a part of surfaces with non-vanishing constant mean curvature (such as sphere, cylinder, and unduloid), biconcave discoid (valid for axisymmetric case), or Willmore surfaces (such as torus, invert catenoid, and so on). Several typically impossible open membranes with free edges are schematically shown in figure~\ref{fig-impopm}. {This theorem suggests that it is very difficult to achieve analytic solutions} to shape equation~(\ref{eq-openm}) with boundary conditions~(\ref{bound1})--(\ref{bound3}) for open lipid membranes. Thus numerical simulations \cite{TuJCP2010,DuLWJCP06,LiJF13} are highly appreciated.

In addition, Tu~\cite{TuJGSP2011} investigated the quasi-exact solution which is defined as a surface with free edges such that the points on that surface exactly satisfy shape equation (\ref{eq-openm}), and most of points (except several discrete points) in the edges abide by boundary conditions~(\ref{bound1})--(\ref{bound3}). Two possible quasi-exact solutions have been achieved: One is a straight stripe cut from a cylindrical surface along the axial direction; another is a twist ribbon which is a part of a minimal surface ($H$=0).

\subsection{Stress tensor of fluid membranes}
{Capovilla \emph{et al.}~\cite{Capovilla2,Capovilla} presented the concept of stress tensor of fluid membranes and then derived the governing equations of open lipid membranes.} We will briefly introduce this key concept based on Helfrich's model and the work by Capovilla \emph{et al.} in this subsection.

\begin{figure}[pth!]
\centerline{\includegraphics[width=5cm]{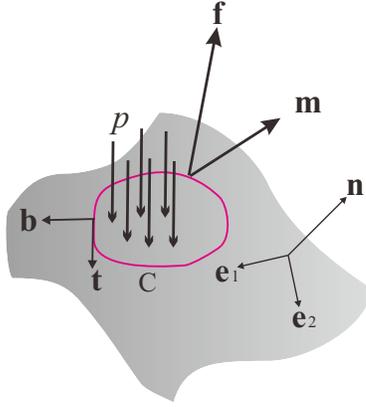}}
\caption{\label{fig-stress}(Color online)
Force balance and moment balance for a domain cut from a lipid membrane.}
\end{figure}

The concept of stress comes from the force balance and the moment balance for any domain in a lipid membrane. As shown in figure~\ref{fig-stress}, we cut a domain bounded by a curve C from the lipid membrane.
At each point, we construct a right-handed orthogonal frame
$\{\mathbf{e}_1,\mathbf{e}_2,\mathbf{n}\}$ with $\mathbf{n}$ being the unit normal vector. A pressure $p$ is
loaded on the surface against the normal direction. $\mathbf{t}$ is the unit tangent vector of curve
C. Unit vector $\mathbf{b}$ is located in the tangent
plane and it is normal to $\mathbf{t}$. Vectors $\mathbf{f}$ and $\mathbf{m}$ represent the force and the moment per unit length applied on curve C by lipids out of the domain,
respectively.

{According to Newtonian mechanics, a physical object is in equilibrium when the force balance and the moment balance are simultaneously satisfied.} It follows that
\begin{eqnarray}
&&\oint_C \mathbf{f}\mathrm{d}s-\int p\mathbf{n}\mathrm{d}A =0,\label{forcb2D01} \\
&&\oint_C \mathbf{m}\mathrm{d}s+\oint_C \mathbf{r}\times
\mathbf{f}\mathrm{d}s-\int \mathbf{r} \times p\mathbf{n}\mathrm{d}A
=0,\label{forcb2D02}
\end{eqnarray}
where $\mathrm{d}s$ and $\mathrm{d}A$ are the arc length element of curve C and the area
element of the domain, respectively. $\mathbf{r}$ represents the position vector of a point on the surface.

We can mathematically define two second order tensors $\mathcal{S}$ and $\mathcal{M}$
such that $\mathcal{S}\cdot\mathbf{b}=\mathbf{f}$ and $\mathcal{M}\cdot\mathbf{b}=\mathbf{m}$.
These two tensors are called stress tensor and bending moment
tensor, respectively. Using the Stokes theorem and considering the arbitrariness of the domain, we can derive the equilibrium equations~\cite{Capovilla2,Tu2008jctn}:
\begin{eqnarray}
&&\mathrm{div~} \mathcal{S} =p\mathbf{n},\label{forcb2D1} \\
&&\mathrm{div~} \mathcal{M} =\mathcal{S}_{1}\times
\mathbf{e}_{1}+\mathcal{S}_{2}\times \mathbf{e}_{2},\label{momb2D1}
\end{eqnarray}
with $\mathcal{S}_1\equiv\mathcal{S}\cdot\mathbf{e}_1$ and
$\mathcal{S}_2\equiv\mathcal{S}\cdot\mathbf{e}_2$. ``$\mathrm{div}$" represents the divergence operator defined on a 2D surface.

As done by Capovilla and Guven~\cite{Capovilla2,Capovilla}, if we consider the Helfrich model, the bending moment tensor and the stress tensor can be express as
\begin{equation}\mathcal{M} =-[k_c(2H+c_0)\mathcal{I}+\bar{k}\mathcal{C}]\times \mathbf{n}\label{eq-momentt}\end{equation}
and
\begin{equation}
\mathcal{S} =[(k_c/2) (2H+c_0)^2+\lambda]\mathcal{I}-k_c (2H+c_0)\mathcal{C}-k_c\mathbf{n}\nabla (2H),\label{eq-stress}
\end{equation}
respectively, where $\nabla$ is the gradient operator defined on a 2D surface. $\mathcal{I}\equiv \mathbf{e}_1\mathbf{e}_1+\mathbf{e}_2\mathbf{e}_2$ represents the 2D unit tensor. $\mathcal{C}\equiv -\nabla \mathbf{n}$ is the curvature tensor whose trace and determinate give the twice of mean curvature ($2H$) and the Gaussian curvature ($K$), respectively. With the consideration of the moment balance equation (\ref{momb2D1}) and the bending moment tensor (\ref{eq-momentt}), the force balance equation (\ref{forcb2D1}) may be further transformed into
\begin{equation}\mathcal{S}\colon \mathcal{C}-2k_c\nabla^2H=p.\end{equation}
Substituting the stress tensor (\ref{eq-stress}) into the above equation, we can readily derive shape equation (\ref{eq-shapeclosed}). Furthermore, governing equations~(\ref{eq-openm})--(\ref{bound3}) of open lipid membranes can also be derived from the stress tensor and the bending moment tensor with the consideration of the Gauss-Bonnet formula and the contribution of line tension~\cite{Capovilla}.

{Particles bound to a soft interface display effective interactions between each other because they deform the shape of the interface. The typical examples are membrane-mediated interactions between  inclusions (e.g. proteins or colloids) adhering to or embedded in lipid membranes~\cite{PincusEPL93,KozlovPRE98,FournierEPL98,OsterBJ98,KoltoverPRL03,Kralchevsky00,MisbahEPJE02,BiscariEPJE02,WeiklEPJE03}.
M\"{u}ller \emph{et al.}~\cite{GuvenEPL05,GuvenPRE05} pointed out that the problem of membrane-mediated interactions between particles may be regarded as a correspondence of general relativity in the low dimension, and that the stress tensor is useful for the investigation of membrane-mediated interactions between particles bound to a lipid membrane.} By using the stress tensor, they expressed the force on a particle as a line integral along any closed contour surrounding the particle~\cite{GuvenEPL05,GuvenPRE05}.

\subsection{Gaussian bending modulus}
The Gaussian bending modulus is significant for the case that the topology of a membrane is changed. For example, it directly influences on the energetics of vesicle divisions. For simplicity, we consider that a spherical vesicle with $R_0$ is divided into two spherical vesicles with identical radius $R_1$. The constraint of constant area requires $R_1 =R_0/\sqrt{2}$. In terms of Helfrich curvature energy (\ref{eq-helfrichcuv}), the bending energy of initial vesicle is $F_i=2\pi k_c (c_0 R_0 -2)^2 +4\pi \bar{k}$. After division, the total bending energy of two spherical vesicles is $F_f=4\pi k_c (c_0 R_0/\sqrt{2}-2)^2+8\pi \bar{k}$. Thus the net gain of energy is
\begin{equation}\Delta F=F_f-F_i=8\pi k_c [1-(\sqrt{2}-1)c_0 R_0]+4\pi \bar{k}.\end{equation}
{This equation implies that the division state (two small spherical vesicles) is more energetically favorable than the initial state (one large spherical vesicle) if $\Delta F<0$, and vice versa.} Hence, the condition for the existence of a spherical vesicle ($\Delta F>0$) may be expressed as
\begin{equation}\bar{k}/k_c >-2 + 2(\sqrt{2}-1)c_0R_0.\end{equation}
Thus it is important for us to know the value of $\bar{k}$. {Theoretical estimation within the framework of liquid crystal theory} implies that $\bar{k}$ may be positive or negative~\cite{OYbook1999}.
One can perturb the local mean curvature and Gaussian curvature with the micropipette technique. However, the integral of gaussian curvature on a surface abides by the famous Gauss-Bonnet formula $\int K \mathrm{d}A=2\pi \chi -\oint\kappa_g \mathrm{d}s$, where $\chi$ depends only on the topology of the surface while $\kappa_g$ is the geodesic curvature of the edge.
Thus the total energy $\bar{k}\int K \mathrm{d}A$ cannot be perturbed with the micropipette technique in conventional experiments, which leads to the difficulty of measuring the value of $\bar{k}$.

To measure the value of $\bar{k}$, we should seek for new ideas, for example, changing the topology of the surface or perturbing the geodesic curvature of the edge. Lorenzen \emph{et al.}~\cite{LorenzenBJ86} regarded a pierced bilayer vesicle as a closed
monolayer vesicle and estimated $\bar{k}\approx -0.83 k_c$. Templer \emph{et al.}~\cite{TemplerLM98} estimated $\bar{k}\approx -0.75 k_c$ from measurements of the swelling behavior in water of inverse bicontinuous cubic
mesophases in a system composed of 1-monoolein, dioleoylphosphatidylcholine, and dioleoylphosphatidylethanolamine. Siegel and Kozlov~\cite{KozlovBJ04} observed the phase behavior of N-mono-methylated dioleoylphosphatidylethanolamine
and determined $\bar{k}\approx -0.83 k_c$. {J\"{u}licher and Lipowsky~\cite{JulicherPRL93} pointed out the possibility to obtain the value of $\bar{k}$ from phase-separated vesicles. Following this idea, Baumgart \emph{et al.}~\cite{JenkinsBJ05} estimated the absolute difference in Gaussian moduli of liquid disorder phase and liquid order phase to be $3.6 k_c$ by fitting the shapes of two-phase vesicles observed in their experiment. Hu \emph{et al.}~\cite{HuJingleiSM11} found that the difference in Gaussian moduli could stabilize the multiple domains of liquid-disorder phase through Monte Carlo simulations.}
Semrau \emph{et al.}~\cite{Semrau08} combined analytical and
experimental approaches to phase-separated vesicles and extracted the value of $\bar{k}\approx-0.31 k_c$. Tu~\cite{TuJCP2010}
fitted the contour line of open lipid membranes observed in the experiment~\cite{Hotani98} by using the theory of open lipid membranes mentioned in section~\ref{sec-openLBs}, and then extracted the value of $\bar{k}\approx-0.12 k_c$.

There are also several estimations from molecular dynamics simulations. By using coarse-grained methods, Brannigan and Brown~\cite{Brown07} estimated $\bar{k}\approx -0.54 k_c$ while den Otter~\cite{denOtter} achieved $\bar{k}\approx -0.03 k_c$. Recently, Hu \emph{et al.}~\cite{Hu-Deserno12,Hu-Deserno13} estimated $\bar{k}\approx -1.0 k_c$ from high accuracy simulations.

If we take the sparse values estimated from experiments and simulations into account, it is still necessary to further investigate the lipid membrane with free edges through tight interplays between theoretical and experimental researches.

\section{Chiral lipid membranes\label{sec-CLM}}
Chiral molecules can form chiral membrane structures~\cite{Nakashima84,Schnur93,Schnur94,Spector98,Spector01}. Fang's group~\cite{Zhaoy05} observed the projected direction of
the DC$_{8,9}PC$ molecules on tubular surfaces and found the 45$^\circ$ departure of direction from the
equator of the tubules at the uniform tilting state. The same group~\cite{Zhaoy06} also observed
lipid tubules with helical ripples. The pitch angles of helical ripples are concentrated on about 5$^\circ$ and 28$^\circ$
\cite{Zhaoy06}. Cholesterol helical stripes with pitch angles
$11^\circ$ and $54^\circ$ were usually observed
\cite{Chung93,Zastavker} in the bile of patients with gallstones. Additionally, Oda \emph{et al.}~\cite{Oda99,Oda02} reported twisted ribbons of achiral cationic amphiphiles interacting
with chiral tartrate counterions. They found
that the width and pitch of twisted ribbons could be tuned by the concentration difference of left- and right-handed tartrate counterions~\cite{Oda99}.
Following the seminal work by Hefrich and Prost~\cite{Helfrich88}, several theoretical models~\cite{oy90,oypra91chm,Nelson92,Selinger93,Selinger96,Selinger01,oy98,tzcpre2007} were developed to explain these experimental results on chiral membranes. In this section, we will briefly review several relevant theoretical achievements.

\subsection{Helfrich-Prost model}
Hefrich and Prost~\cite{Helfrich88} assumed that the chiral molecules stay in the Smectic C$^*$ phase at which the direction of the
molecules is tilted from the normal of membranes at a constant
angle. Select a locally right-handed orthogonal frame $\{\mathbf{n}, \mathbf{m}, \mathbf{p}\}$, where $\mathbf{n}$ is the normal vector of the membrane, $\mathbf{m}$ denotes the projection of tilting direction on the membrane, and $\mathbf{p}$ coincides with the axis of the ferroelectric polarization. According to the symmetry argument~\cite{Helfrich88}, the bending energy per unit area may be expressed as the sum of a complete set of independent invariants of quadratic
and linear order in $\nabla\mathbf{n}$, which reads
\begin{eqnarray}\mathcal{E}_\mathrm{ch}&=&(1/2)k_{mm}(\mathbf{m}\cdot\nabla\mathbf{n}\cdot\mathbf{m})^2+(1/2)k_{pp}(\mathbf{p}\cdot\nabla\mathbf{n}\cdot\mathbf{p})^2\nonumber\\
&+&k_{mp}(\mathbf{m}\cdot\nabla\mathbf{n}\cdot\mathbf{p})^2+(1/2)\bar{k}[(\mathrm{Tr}\nabla\mathbf{n})^2-\mathrm{Tr}(\nabla\mathbf{n})^2]\nonumber\\
&-&k_{m}(\mathbf{m}\cdot\nabla\mathbf{n}\cdot\mathbf{m})-k_{p}(\mathbf{p}\cdot\nabla\mathbf{n}\cdot\mathbf{p})-h(\mathbf{m}\cdot\nabla\mathbf{n}\cdot\mathbf{p}),\label{eq-HelfCLM}
\end{eqnarray}
where the operator ``$\mathrm{Tr}$" represents the trace of a tensor.
The first three terms in the above equation reflect the anisotropic bending. The fourth term represents the contribution of Gaussian curvature. The first two linear terms lead to two spontaneous curvatures $c_{0m}=k_m/k_{mm}$ and $c_{0p}=k_p/k_{pp}$. The last term represents the effect of molecular chirality, which vanishes for achiral membranes because $\mathbf{p}$ and $-\mathbf{p}$ are physically
equivalent for the achiral membranes. For completeness, with the consideration of orientational order, the terms $(\mathbf{p}\cdot\nabla\mathbf{m}\cdot\mathbf{p})^2$, $(\mathbf{m}\cdot\nabla\mathbf{m}\cdot\mathbf{p})^2$,
$(\mathbf{p}\cdot\nabla\mathbf{m}\cdot\mathbf{p})$, and $(\mathbf{m}\cdot\nabla\mathbf{m}\cdot\mathbf{p})$ may be added in equation~(\ref{eq-HelfCLM}).

For simplicity, Hefrich and Prost discussed the special case of an isotropic
bending by taking $k_{mm}=k_{mp}=k_{pp}=k_c$. Equation (\ref{eq-HelfCLM}) may be written as $\mathcal{E}_\mathrm{ch}=(k_c/2)(\mathrm{Tr}\nabla\mathbf{n})^2-h(\mathbf{m}\cdot\nabla\mathbf{n}\cdot\mathbf{p})$.
For the uniform tilting state in a cylindrical membrane with radius $R$, the elastic energy density may be further written as $\mathcal{E}_\mathrm{ch}=(k_c/2R^2)-(h/2R)\sin2\varphi$, where $\varphi$ represents the angle between the titling direction and the circumferential direction of the cylinder.
Obviously, this energy takes minimum at $\varphi=\pi/4$ for $h>0$ and $\varphi=-\pi/4$ for $h<0$, which may provide a good explanation to the experimental facts observed in Ref.~\cite{Zhaoy05}. If the bending rigidity is anisotropic, the angle $\varphi$ of minimum energy needs not be $\pi/4$. The observation in Ref.~\cite{Chung93,Zastavker}
might correspond to this case.

Based on the Hefrich-Prost model, Ou-Yang and Liu~\cite{oy90,oypra91chm} explained the transition sequence
from the vesicle to twisted ribbon then to helical stripe observed in experiment~\cite{Nakashima84}. In particular, they also found that the chiral term could be further expressed as \begin{equation}f_\mathrm{ch}=-h\tau_{\mathbf{m}},\label{chiralen}\end{equation}
where $\tau_{\mathbf{m}}$ is the
geodesic torsion along $\mathbf{m}$, the projection vector of titling direction on the membrane.

Nelson and Powers~\cite{Nelson92} adopted the Hefrich-Prost model to investigate a rigid chiral membrane with an assumption that the bending rigidities $k_{mm}$, $k_{mp}$, $k_{pp}$ are much larger than $k_BT$ while the chiral coupling parameter $h$ is relative small. By using the renormalization group theory, they showed how thermal fluctuations could reduce the effective value of the chiral coupling constant.

\subsection{Selinger-Schnur model}
Selinger and Schnur~\cite{Selinger93} refined and developed the Helfrich-Prost model for chiral lipid tubules. The free energy in their model contains three types of contributions.
The first one is the curvature free energy
\begin{equation}F_\mathrm{curv}=\int \mathrm{d}A [(k_c/2)(1/R)^2],\label{eq-curvetb}\end{equation}
where $R$ is the radius of a tubule while $k_c$ is the bending rigidity.
The second term is the tilting free energy
\begin{equation}F_\mathrm{tilt}=\int \mathrm{d}A [(-a/2)\theta^2 +(b/4)\theta^4],\end{equation}
where $\theta$ represents the angle between the direction of a lipid molecule and the normal direction of the tubule at the position of that lipid molecule. This free energy has the Landau-like form: $a=\alpha (T_c -T)$ and $b>0$ are two Landau coefficients. $T$ and $T_c$ represent the temperature of environment and the critical temperature, respectively. This free energy can describe the transition from the tilting phase to the untilting phase when the temperature is increased.
The third term is the {Frank free energy~\cite{Frank58}} due to the distortions of direction field arranged by lipid molecules:
\begin{equation}F_\mathrm{Frank}=\int \mathrm{d}A[(k_1/2)(\nabla_3\cdot \mathbf{l})^2 +(k_2/2)(\mathbf{l}\cdot\nabla_3 \times\mathbf{l}-q_0)^2+(k_3/2)(\mathbf{l}\times\nabla_3 \times\mathbf{l})^2],\label{eq-frank}\end{equation}
where $k_1$, $k_2$ and $k_3$ are the elastic constants for splay,
twist, and bend distortions, respectively. $\nabla_3 $ represents the gradient operator in a 3D Euclidean space. The parameter $q_0$ represents the
chirality of lipid molecules. The unit vector $\mathbf{l}$ represents the direction of
each lipid molecule.

\begin{figure}[pth!]
\centerline{\includegraphics[width=6cm]{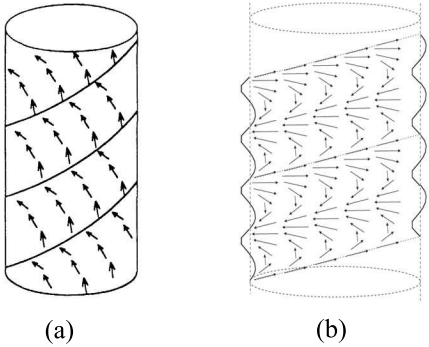}}
\caption{\label{fig-module}Tubules: (a) with helically modulated tilting state~\cite{Selinger93}; (b) with ripples.}
\end{figure}

The total free energy may be expressed as $F_\mathrm{total}=F_\mathrm{curv}+F_\mathrm{tilt}+F_\mathrm{Frank}$. Adopting this free energy, Selinger and Schnur~\cite{Selinger93} predicted a tubule with helically modulated tilting state. As shown in figure~\ref{fig-module}a, several helical stripes form the tubule while the tilting direction of molecules in each helical stripe is invariant along the direction of the helix.
Next, Selinger \emph{et al.}~\cite{Selinger96} generalized their previous model to describe chiral lipid membranes rather than the perfect cylindrical tubules. They predicted an imperfect tubule with helical ripples as shown in figure~\ref{fig-module}b.
Then, they proposed a scenario for the kinetic
evolution from flat membranes (or large vesicles) into
tubules. When a flat membrane is
cooled from an untilting phase into a tilting phase, the tilting
order emerges. The tilted chiral molecules form
a series of stripes separated by domain walls. Each stripe then forms a helix with ripples. These helices may grow wider and wider to form a tubule with helical ripples as shown in figure~\ref{fig-module}b.

Komura and Ou-Yang~\cite{oy98} argued that the Frank free energy (\ref{eq-frank}) has implicitly contained the curvature free energy (\ref{eq-curvetb}). They merely began with the Frank free energy and considered two classes of helical stripes: one is called P-helix which is at the uniform tilting phase such that the molecules nearby the domain walls are parallel packing; the other is called A-helix which is at the modulated tilting state such that the molecules nearby the domain walls are antiparallel packing. They found that the A-helix can explain the helical stripes with low-pitch angle $11^{\circ}$ observed in the experiment~\cite{Chung93} while the P-helix exactly corresponds to the helical stripes with high-pitch angle $54^{\circ}$ observed in the experiment~\cite{Chung93}.

\subsection{Concise theory of chiral lipid membranes}
{Due to the complicated form of the free energy used in above theories
\cite{Helfrich88,oy90,oypra91chm,Nelson92,Selinger93,Selinger96,Selinger01,oy98},
the general Euler-Lagrange equations corresponding to the free energy are expected to be so intricate that they have not been explicitly written out in the previous work.} Thus no one has unambiguously judged whether a configuration is a genuinely equilibrium structure or not. Tu and Seifert~\cite{tzcpre2007} constructed a simplified theory of chiral lipid membranes which could overcome this difficulty to some extent.

\begin{figure}[pth!]
\centerline{\includegraphics[width=6cm]{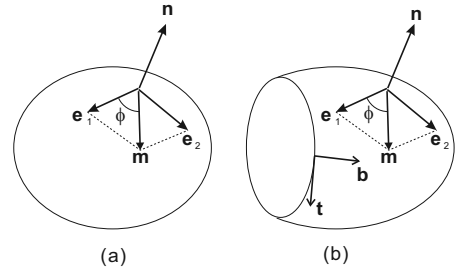}}
\caption{\label{fig-chmb}
Representations of chiral lipid membranes. $\{\mathbf{e}_1,\mathbf{e}_2,\mathbf{n}\}$ is a right-handed orthogonal frame with $\mathbf{n}$ being the normal vector of membrane surface. $\mathbf{m}$ represents the unit project vector of a lipid molecule on the membrane surface. (a) A vesicle is described by a closed smooth surface. (b) An open membrane is represented by a smooth surface with a boundary curve. $\mathbf{t}$ is the tangent vector of the boundary curve, and
$\mathbf{b}$, in the tangent plane of the surface, is perpendicular
to $\mathbf{t}$.}
\end{figure}

Chiral lipid membranes can be represented as smooth surfaces with or without boundary curves as shown in figure~\ref{fig-chmb}. Tu and Seifert~\cite{tzcpre2007} assumed that the free energy per unit area contains three contributions: The first one is the isotropic curvature energy, which is taken as the Helfrich form (\ref{eq-helfrichcuv}). The second one comes from the chirality of molecules, which is taken as equation~(\ref{chiralen}). Without loss of generality, $h$ is assumed to be positive. The third one comes from the orientational variation of tilting order~\cite{Nelson87} in the membrane surface, which is taken as
$f_\mathrm{tilt}=(k_{f}/2)[(\mathrm{curl~} \mathbf{m})^{2}+(\mathrm{div~}\mathbf{m})^{2}]$,
where $k_f$ is an elastic constant. Note that ``curl" is the curl operator defined on a 2D surface, which gives an scalar when it operates on a vector. Introducing a spin connection field $\mathbf{S}$ satisfying
$\mathrm{curl~}\mathbf{S}=K$, one can derive $(\mathrm{curl~} \mathbf{m})^{2}+(\mathrm{div~}\mathbf{m})^{2}=(\nabla\phi-\mathbf{S})^{2}$
through simple calculations~\cite{Nelson87}, where $\phi$ is the angle between vector $\mathbf{m}$ and the base vector $\mathbf{e}_1$. Thus, the free energy per unit area may be expressed as
the following concise form:
\begin{equation}
\mathcal{E}=\frac{k_{c}}{2}(2H+c_{0})^{2}+\bar{k}K-h\tau_{\mathbf{m}}+\frac{k_{f}}{2}\mathbf{v}
^{2},\label{energy2}
\end{equation}
with $\mathbf{v}\equiv\nabla\phi-\mathbf{S}$. From a symmetric point of view, this special form
is the minimal construction
including the bending, the chirality, and the tilting order for given
vector field $\mathbf{m}$ on the membrane surface and normal vector field $\mathbf{n}$ of membrane surface. In recent work, Napoli and Vergori~\cite{Napoli12,Napoli13} derived the effective 2D free energies for chiral lipid membranes from the 3D Frank theory~\cite{Frank58} for cholesteric liquid crystals. Their work implies that free energy density (\ref{energy2}) is available for the strongly twisted cholesterics while an additional term should be included in equation~(\ref{energy2}) for the weakly twisted cholesterics.

The free energy for a closed chiral lipid vesicle may be expressed as
\begin{equation}F=\int \mathcal{E}\mathrm{d} A +\lambda A+p V,\label{closedFE}\end{equation}
where $A$ is the area of the membrane and $V$ is the volume enclosed by the vesicle. $\lambda$ and $p$ are two Lagrange multipliers to implement area and volume constraints.
Using the surface variational method~\cite{TuJPA04}, Tu and Seifert~\cite{tzcpre2007} derived two
governing equations for equilibrium configurations:
\begin{equation}
2\tilde{h}(\kappa_{\mathbf{m}}-H)-\tilde{k}_{f}(\nabla^{2}\phi-\mathrm{div~}\mathbf{S})=0 \label{EL1}
\end{equation}
and
\begin{eqnarray}
&&\hspace{-0.5cm}2\nabla^{2}H+( 2H+c_{0}) (
2H^{2}-c_{0}H-2K)-2\tilde{\lambda} H+\tilde{p}\nonumber\\
&&\hspace{-0.5cm}+\tilde{h}[  \mathrm{div~}( \mathbf{m}\mathrm{~curl~}\mathbf{m})
+\mathrm{curl~}(
\mathbf{m}\mathrm{~div~}\mathbf{m})]\nonumber\\
&&\hspace{-0.5cm}+\tilde{k}_{f}[( \kappa_{\mathbf{v}}-H) \mathbf{v}^{2}+
\nabla\mathbf{v}\colon \mathcal{C}] =0\label{EL2}
\end{eqnarray}
with reduced parameters $\tilde{h}\equiv h/k_c$, $\tilde{k}_{f}\equiv {k}_{f}/k_c$, $\tilde{p}\equiv p/k_c$, and $\tilde{\lambda}\equiv \lambda/k_c$. $\kappa_{\mathbf{m}}$ and $\kappa_{\mathbf{v}}$ are the normal
curvatures along the directions of vectors $\mathbf{m}$ and $\mathbf{v}$,
respectively. Note that Tu and Seifert~\cite{tzcpre2007} did not consider singular points for closed vesicles. Generally, the defects should have some effects on the morphology of vesicles. Jiang \emph{et al.}~\cite{JiangHPRE07} found that the tilting order significantly influences on the shapes of chiral lipid membranes with narrow necks.
Xing \emph{et al.}~\cite{Xingpnas12,Xingprl08} also found that inevitable topological defects in chiral lipid vesicles with spherical topology play essential roles in controlling the final morphology of vesicles.

The free energy of a chiral lipid membrane with a free edge may be expressed as
\begin{equation}F=\int  \mathcal{E} \mathrm{d} A+\lambda A +\gamma L ,\label{openFE}\end{equation}
where $A$ is the area of the membrane and $L$ the total length of the edge. $\gamma$ represents the line tension of the edge. Tu and Seifert~\cite{tzcpre2007} found that the governing equations are the same as equations~(\ref{EL1}) and (\ref{EL2}) with vanishing $\tilde{p}$, simultaneously, the following boundary conditions should be imposed on the free edge:
\begin{eqnarray}&&v_{b}=0,\label{BC1}\\
&&(1/2)(2H+c_{0})^{2}+\tilde{k}K-\tilde{h}\tau_{\mathbf{m}}+(\tilde{k}_{f}/2)\mathbf{v}^2+\tilde{\lambda}+\tilde{\gamma}\kappa_{g}=0,\label{BC2}\\
&&(2H+c_{0})+\tilde{k}\kappa_{n}-(\tilde{h}/2)\sin2\bar{\phi}=0,\label{BC3}\\
&&\tilde{\gamma}\kappa_{n}+\tilde{k}\dot{\tau}_{g}-2\partial H/\partial\mathbf{b}-\tilde{h}(v_{t}+\dot{\bar{\phi}})\sin2\bar{\phi}
+\tilde{k}_{f}\kappa_{n}v_{t}=0,\label{BC4}\end{eqnarray} with
$v_b\equiv \mathbf{v}\cdot \mathbf{b}$ and $v_t\equiv \mathbf{v}\cdot \mathbf{t}$.
$\kappa_{n}$, $\tau_{g}$ and $\kappa_{g}$ represent the normal curvature,
the geodesic torsion, and the geodesic curvature of the boundary curve
(i.e., the edge), respectively. The `dot' represents the derivative with
respect to arc length parameter $s$. $\bar\phi$ is the angle between $\mathbf{m}$ and
$\mathbf{t}$ at the boundary curve. Boundary conditions (\ref{BC1})--(\ref{BC4}) are also available for a chiral lipid membrane with
several edges since they describe the force balance and the moment balance
in the edge.

The concise theory mentioned above is consistent with the previous experiments \cite{Schnur94,Spector98,Zhaoy05} on self-assembled chiral lipid membranes of DC$_{8,9}$PC.
This theory does not permit
genuinely helical stripes with free edges in a uniform tilting
state. It also does not admit tubules with helically modulated tilting state which are
energetically less favorable than tubules with helical ripples.
Up to the first order perturbation, Tu and Seifert~\cite{tzcpre2007} estimated the pitch
angles of helical ripples to be about
0$^\circ$ and 35$^\circ$, which are close to the most frequent
values 5$^\circ$ and 28$^\circ$ observed in the experiment~\cite{Zhaoy06}.

\begin{figure}[pth!]
\centerline{\includegraphics[width=9cm]{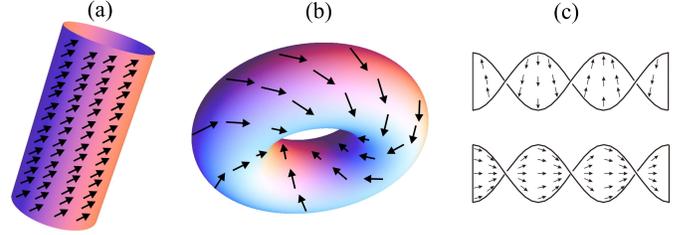}}
\caption{\label{fig-solutchm}(Color online) Three analytic solutions to the governing equations of chiral lipid membranes.}
\end{figure}

Three analytic solutions~\cite{tzcpre2007} to equations~(\ref{EL1}) and (\ref{EL2}) for chiral lipid vesicles or those with boundary conditions (\ref{BC1})--(\ref{BC4}) for open chiral membranes were obtained. The first one is a perfect tubule with uniform tilting state as shown in figure~\ref{fig-solutchm}a where
the projected direction of the molecules on the tubular surface departs 45$^\circ$ from the equator of the tubule. The second one is a torus with uniform tilting state as shown in figure~\ref{fig-solutchm}b where
the projected direction of the molecules on the toroidal surface departs 45$^\circ$ from the equator of the torus. Interestingly, the ratio of two generation radii of the torus may be expressed as
\begin{equation}\varrho=\sqrt{(2-\tilde{k}_f)/(1-\tilde{k}_f)}.\label{eq-torusrad2}\end{equation}
Different from lipid torus mentioned in section \ref{sec-Lipshpcls}, this ratio can be larger than $\sqrt{2}$ for non-vanishing $\tilde{k}_f$. The third one corresponds to twisted ribbons as shown in figure~\ref{fig-solutchm}c. The twisted ribbon in the top of figure~\ref{fig-solutchm}c is left-handed and the projected direction of the molecules is perpendicular to the edge. On the contrary, the twisted ribbon in the bottom of figure~\ref{fig-solutchm}c is right-handed and the projected direction of the molecules is parallel to the edge. The ratio of the width to the pitch of twisted ribbons is predicted to be proportional to the relative
concentration difference of left- and right-handed enantiomers, which is in good
agreement with the experiment~\cite{Oda99}.

\section{Conclusions}
Since Helfrich's seminal work~\cite{helfrich73} was published in 1973, great achievements have been made in the field of elasticity of membranes during the past forty years. Helfrich's successors such as Prost, Liopwsky, Ou-Yang, Seifert, Selinger, Guven, Deserno, and so on, have made significant contributions in theoretical aspect during the development of elasticity of membranes. In this review, according to our personal prospect, we have reported several theoretical advances achieved by these researchers following the Helfrich theory of fluid membranes. We have presented the governing equations describing equilibrium configurations of lipid vesicles, lipid membranes with free edges, and chiral lipid membranes. We have also provided several special solutions to these equations and their corresponding configurations.

Although such great progress has been made, there still remain several challenges in theoretical aspect.

(i) Analytic solutions to shape equation (\ref{eq-shapeclosed}). Can we further find analytic solutions to the shape equation rather than the sphere, the torus and the biconcave discoid such that they correspond to closed vesicles without self-contact or self-intersection?

(ii) Minimal geodesic disk~\cite{Tucpb2013}. Although we have introduced the theorem of non-existence in section \ref{sec-openLBs},  we cannot exclude a possible nontrivial solution: a disk on some minimal surface whose boundary curve has constant geodesic curvature but vanishing normal curvature. This kind of disks is briefly called minimal geodesic disk. A flat circular disk is a trivially minimal geodesic disk. Can we find a nontrivially minimal geodesic disk rather than the flat one? We conjecture that
the flat circular disk is the unique minimal geodesic disk. An argument like this conjecture has been implicated in Almgren's proof of the general codimension sharp isoperimetric inequality~\cite{Morganbook}. Whether this conjecture is true or false is still an open mathematical question~\cite{Fmorgan}.

(iii) {Lipid vesicles with multi-domains. Separation of liquid phases and formation of domains in giant vesicles of ternary mixtures of phospholipids and cholesterol were experimentally observed~\cite{JenkinsBJ05,Baumgart03,Veatch03,GudhetiBJ11}.} Although lipid vesicles with two or several domains have been investigated from theoretical aspects \cite{TuJPA04,Tu2008jctn,JulicherPRL93,JenkinsBJ05,GozdzPRE99,JuelicherPRE96,Wangdu08,Das2009}, there is still a lack of rigorously analytic solutions to the general governing equations~\cite{TuJPA04,Tu2008jctn} describing the configurations of lipid vesicles with multi-domains.

\section*{Acknowledgements}
This review is dedicated to the 80th birthday of Prof. Dr. Wolfgang Helfrich. ZCOY is sincerely grateful to Helfrich for his valuable advice and kind collaborations from 1987 to 1990. ZCT thanks Pan Yang and Yang Wang for their carefully proofreading the manuscript. The authors are also grateful to finical supports from the National Natural Science Foundation of China (Grant Nos. 11274046 and 11322543).


\end{document}